# Synthesis of a 3D network of Pt nanowires by atomic layer deposition on carbonaceous template

Shaoren Deng, ‡[a] Mert Kurttepeli, ‡[b] Stella Deheryan, [c] Daire J. Cott, [c] Philippe M. Vereecken, [c, d] Johan A. Martens, [d] Sara Bals, [b] Gustaaf van Tendeloo, [b] and Christophe Detavernier*[a]

[a]Department of Solid State Sciences, Ghent University, Krijgslaan 281/S1, B-9000 Ghent, Belgium

[b]Electron Microscopy for Materials Science (EMAT), University of Antwerp, Groenenborgerlaan 171, B-2020, Antwerp, Belgium

[c]IMEC, Kapeldreef 75, B-3001 Leuven, Belgium

[d]Center for Surface Chemistry and Catalysis, KU Leuven, Kasteelpark Arenberg 23, B-3001 Heverlee, Belgium

‡ These authors contributed equally.

## Abstract

The formation of a 3D network composed of free standing and interconnected Pt nanowires is achieved by a two-step method, consisting of conformal deposition of Pt by atomic layer deposition (ALD) on a forest of carbon nanotubes and subsequent removal of the carbonaceous template. Detailed characterization of this novel 3D nanostructure was carried out by transmission electron microscopy (TEM) and electrochemical impedance spectroscopy (EIS). These characterizations showed that this pure 3D nanostructure of platinum is self-supported and offers an enhancement of the electrochemically active surface area by a factor of 50.

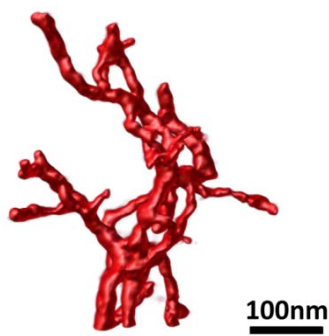

## Introduction

As an effective catalyst, platinum (Pt) has been widely used in many applications including water splitting, fuel cells and automobile emission control.[1-3] Different synthesis methods have been so far employed to synthesize Pt nanostructures with various morphologies. Electrospinning was for instance successfully carried out for the synthesis of Pt nanowires.[4] Yang et al. reported the production of Pt nanoparticles by reducing chloroplatinic acid ($H_2PtCl_6$) in formic acid at low temperature.[5] In case of porous Pt structures synthesis, the use of a template material with a large surface area is widely preferred. Sun et al. used an aqueous method to grow ultra-small Pt nanowires on nitrogen-doped carbon nanotubes.[6] Sattayasamitsathit et al. demonstrated a delicate three-dimensional (3D) porous carbon structure decorated with Pt nanoparticles by electro-depositing.[7] 3D ordered networks of platinum nanowires were synthesized by S. Akbar et al. using lipid as an electrochemical template with a high surface area. Dealloying Pt-M (Al, Fe, Au…) alloy is another popular way to fabricate porous Pt nanowires and has therefore been under intensive research in the recent years.[8-10]

Atomic layer deposition (ALD) is an effective method to functionalize porous materials and tune the pore size.[11-12] Different types of porous materials, e.g. Pt nanotubes or hybrid catalysts, have been successfully produced using ALD.[13-15] Liu et al. for instance deposited Pt nanoparticles on carbon nanotubes (CNTs) support for fuel cell applications.[16] On the other hand, support-free and porous Pt nanostructures can be advantageous for certain applications, e.g., in aggressive environments where the supports are potentially not stable. As an example, Pt nanoparticles dispersed on a carbon support may suffer from instability issues during oxygen reduction reactions.[17-18]

Here, we report ALD of Pt on ordered multi-walled carbon nanotubes (MWCNTs) and the formation of a 3D network comprising pure self-supporting Pt nanowires upon removal of the MWCNTs through annealing. Electrochemical impedance spectroscopy (EIS) tests showed that the resulting 3D network of Pt nanowires exhibit a large and electrochemically active surface area. Henceforth, the ALD Pt coated MWCNTs and the 3D network of Pt nanowires (i.e. after removal of the carbonaceous support) will be referred to as "Pt-MWCNTs" and "3D-Pt" throughout this paper, respectively.

**Experimental**

MWCNTs with 10μm height were synthesized using a $C_2H_4/H_2$ mixture gas in a microwave (2.45 GHz) plasma enhanced chemical vapor deposition chamber (PECVD, TEL, Japan) on 1nm Co/70nm TiN coated Si wafers. A home built ALD reactor with base pressure of ~$5*10^{-7}$ mbar was used for ALD depositions.[19] ALD of Pt was carried out at 200°C with alternating pulses of (methylcyclopentadienyl) trimethylplatinum

(MeCpPtMe$_3$) (Sigma Aldrich) and ozone with a concentration of 150 µg per mL supplied by an ozone generator (Yanco Industry LTD).[20] For each pulse, the pulse pressure was ~0.8 mbar. A Bruker D8 Discovery system was used for the X-ray fluorescence (XRF) measurements. Scanning electron microscopy (SEM) and energy dispersive X-ray spectroscopy (EDS) measurements were done in FEI Quanta 200F. The in situ X-ray Diffraction (XRD) measurements and the post deposition annealing were carried out in a home modified Bruker D8 Discovery system. Transmission electron microscopy (TEM) samples were prepared by scraping off the Pt-containing layer from the substrate surface, and suspending the resulting powder in ethanol. A drop of this suspension was deposited on a carbon coated TEM grid. Bright-field TEM (BFTEM) images and selected area electron diffraction (SAED) patterns were acquired using a FEI Tecnai F20 TEM operated at 200 kV. High-angle annular dark field scanning TEM (HAADF-STEM) images and energy-dispersive X-ray (EDX) elemental maps were collected using a FEI Titan 60-300 TEM operated at 300 kV, equipped with the ChemiSTEM system for EDX analysis. HAADF-STEM electron tomography tilt series were collected using either a FEI Titan 60-300 TEM operated at 300 kV, or using a FEI Tecnai F20 TEM operated at 200 kV, both of which equipped with an advanced tomography holder from Fischione Instruments and the FEI XPlore3D software package. For the tomography experiments, 71 HAADF-STEM images were acquired over the range of ±70° with 2° tilt increments. Alignment and reconstruction of the tomography data were carried out using the FEI Inspect3D software package. The reconstructions were performed using the "Simultaneous Iterative Reconstruction Technique" (SIRT) with 25 iterations implemented in Inspect 3D. Amira software (Visage Imaging GmbH) was used for visualizations. Electrochemical characterization of the samples was done with 1M Na$_2$SO$_4$ in H$_2$O under ambient conditions and a room temperature. All the chemicals were delivered from Sigma-Aldrich, USA. The experiments were performed in a three-electrode electrochemical cell, with a Pt mesh (Sigma-Aldrich, US) as counter electrode and a Ag/AgCl, KCl (sat'd) (BASi USA) as reference electrode. Top contacts were made using a copper tape (Sigma-Aldrich, U.S.A) on four corners of the working electrode. The active surface area in both set-ups was equal to 0.95 cm$^2$. Measurements were done using an Autolab PGSTAT302N potentiostat integrated with frequency response analyzer and controlled by Nova 1.8 software (Ecochemie, Netherlands). All electrochemical impedance Spectroscopy Measurements were performed after Cyclic Voltametry measurements on the same sample. CV measurements with four consecutive scans with a sweep rate of 20mV/s from negative voltages towards positive were carried out. EIS measurements were performed over a frequency range of 10 KHz to 1 Hz with an AC amplitude of 10 mV without agiation. Nova V1.8 software was used to do nonlinear least-squares fitting of the EIS spectra.

**Results and discussion**

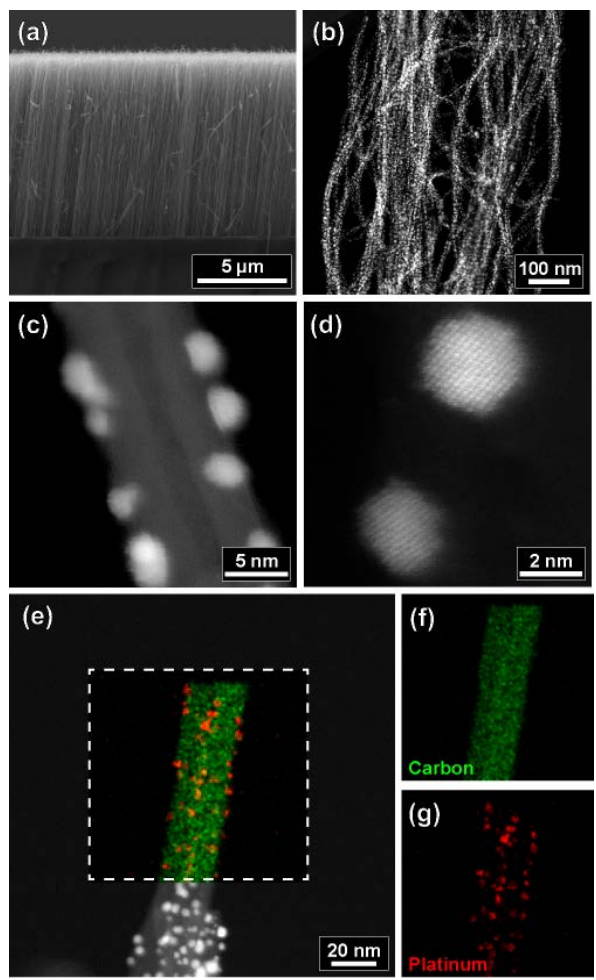

**Figure 1** (a) and (b) are the SEM and HAADF-STEM images of the as deposited 50 cycles Pt-MWCNTs sample. (c) and (d) are the high resolution HAADF-STEM images showing crystalline Pt nanoparticles coated on a single MWCNT. (e), (f) and (g) show the STEM/EDX analysis on the 50 cycles Pt-MWCNTs sample. Color-mixed elemental map of (f) carbon (green) and (g) platinum (red) is embedded on the (e) HAADF-STEM image.

Conformal coating of MWCNTs requires adequate pulse times of precursor and oxidant during the ALD process. In situ and ex situ X-ray fluorescence (XRF) characterization has been proved to be an effective method for identifying the saturation condition for ALD coating of porous materials.[21] 50 cycles of ALD were applied on MWCNTs with different pulse times of 20, 30, 40 and 60 sec of MeCpPtMe$_3$ and ozone with a fixed pumping time of 60 sec for each half cycle. Ex situ XRF measurement on this batch of samples showed different intensities. For pulse times of 40 sec and longer, the intensity of the PtL fluorescence leveled, which indicates that 40 sec exposure is sufficient for the conformal coating on the MWCNTs (Fig. S1). A scanning electron microscopy (SEM) image of the Pt-MWCNTs (Fig. 1(a)) shows that after 50 cycles Pt coating, no obvious morphological changes were observed at the micrometer-scale after ALD. A TEM study of the samples (See Fig. 1(b) and (c)) revealed that Pt-ALD on the MWCNTs results in the formation of nanoparticles attached to the MWCNTs. The high-resolution HAADF-STEM image given in Fig. 1 (d) illustrates the crystallinity of the as deposited Pt nanoparticles. In order to investigate the purity of the Pt nanoparticles, EDX elemental maps were collected from a region specified on the HAADF-STEM image, as shown in Fig. 1 (e). The color-coded elemental map indicates the carbon and platinum content of the specified area and proves that the Pt nanoparticles are indeed attached to the MWCNTs. In order to characterize the Pt-coated MWCNTs in 3D, an electron tomography experiment was performed. The visualization of the 3D reconstructed volume is given in Fig. 2. An animated version of the tomogram is provided in the supporting information as a video (See supporting information). The formation of randomly distributed Pt nanoparticles on the MWCNTs is clearly observed in the 3D visualization. The size distribution of the Pt nanoparticles was obtained from the electron tomography experiment. It was found that the nanoparticles have an average particle radius around 3nm, as shown in Fig. 2.

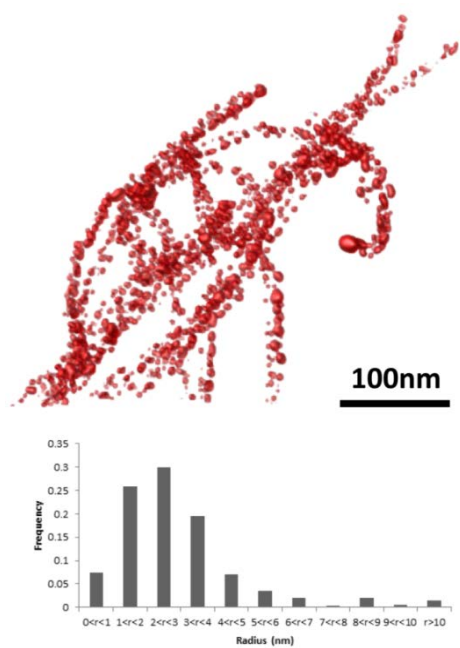

**Figure 2** The visualization of the 3D tomography and the distribution of Pt nanoparticle size from 50 cycles Pt-MWCNTs sample (only Pt nanoparticles are shown).

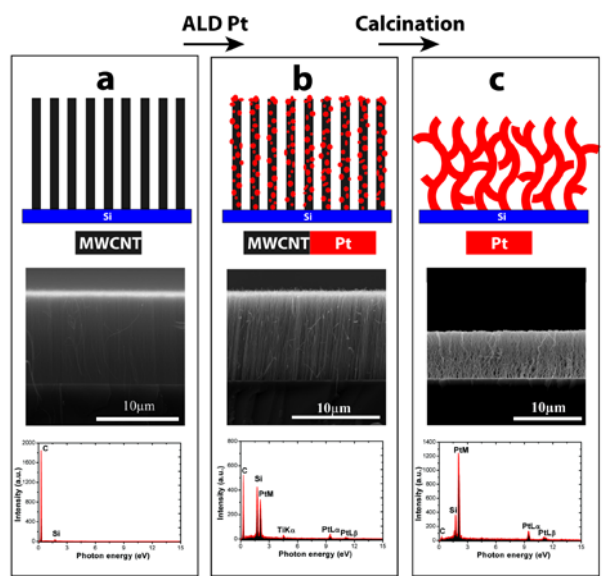

**Figure 3** A combined figure of schematic illustration, SEM images and EDS results shows the morphology and elementary transformation from MWCNTs (a) to Pt-MWCNTs (b) and to 3D-Pt (c).

In view of the wide spacing in between neighboring Pt nanoparticles on the MWCNTs, one could expect that removal of the carbonaceous support would induce a complete collapse of the structure. Surprisingly, the overall forest-type morphology was preserved after calcination of the Pt-MWCNTs. In Fig. 3, SEM images and EDS results are used to schematically illustrate the transformation of Pt-MWCNTs into a free standing 3D-Pt nanostructure. Even though the MWCNTs were removed during the calcination, the overall structure was largely preserved; although a clear shrinkage of the height can be observed. Detailed characterization of the resulting 3D-Pt sample was carried out by TEM (see Fig. 4.). Fig. 4(a) and (b) depict this 3D network comprising Pt nanowires with different diameter ranging from 10 to 50nm. Such variation of the diameter is likely to result from the merging of Pt coating on adjacent MWCNTs into larger crystalline wires, whereas the Pt coating on isolated MWCNTs forms thinner nanowires. The high-resolution HAADF-STEM image shown in Fig. 4 (c) reveals that these Pt nanowires are zigzag shaped and composed of crystalline nanoparticles. The SAED pattern (see Fig. 4 (d)) shows no diffraction rings from the graphite phase but a clear appearance of diffraction rings from the Pt, confirming the removal of the MWCNTs subsequent to annealing. The 3D nature of the network of Pt nanowires is revealed by HAADF-STEM electron tomography (see Fig. 5). An animated version of the tomogram is provided in the supporting information as a video. Unlike the as-deposited sample where Pt nanoparticles were observed to be attached to MWCNTs (see Fig. 2), the annealed sample exhibits a 3D Pt network consisting of many interconnected Pt nanowires with different diameters. Necks throughout all the Pt nanowires were observed, which could be due to the migration and agglomeration of small Pt nanoparticles, and therefore resulting in thinner regions. From the HAADF-STEM electron tomography of the 50 cycles Pt-MWCNTs sample, the so-called "orthoslices" were collected from specified locations at the visualization. These orthoslices also show a similar feature, that all the MWCNTs are removed completely and solid Pt nanowires are formed (Fig. S2).

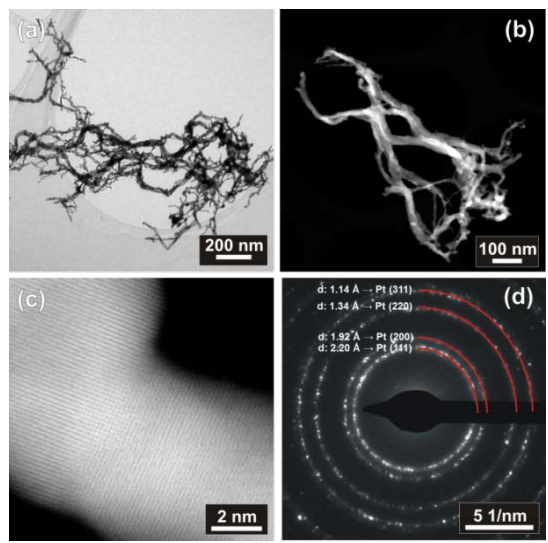

**Figure 4** (a) and (b) are BF-TEM and HAADF-STEM images of 50 cycles 3D-Pt sample. (c) is the high resolution HAADF-STEM images of 50 cycles 3D-Pt sample. (d) shows the SAED pattern of the part of 50 cycles 3D-Pt sample given in (a).

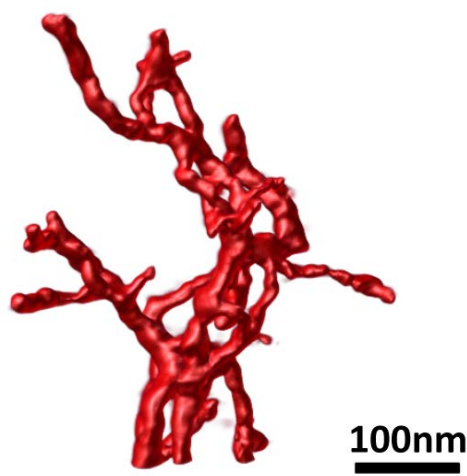

**Figure 5** 3D tomography picture of a section cut from 50 cycles 3D-Pt sample.

A detailed investigation of the carbon removal was carried out by EDS on several quenches of 50 cycles Pt-MWCNTs sample and in situ XRD during the isothermal process, as shown in Fig. 6 (a) and (b). For as grown MWCNTs, the intensity of the carbon signal slightly decreases when the annealing temperature reaches 550°C and decreases significantly from 550 to 650°C, which indicates that the oxidation temperature is around 600°C for the uncoated MWCNTs. For the 50 cycles Pt-MWCNTs, the carbon EDX signal exhibits a similar evolution. However, the drastic descent of the carbon signal starts at a temperature between 350 and 400°C. This implies that the oxidation temperature of the MWCNTs has been significantly reduced due to the Pt nanoparticle coating on the MWCNTs. This might be attributed to the catalytic effect of the Pt nanoparticles, which seems to facilitate the oxidation process of the MWCNTs.[23] Since the 50 cycles Pt-MWCNTs sample has an uncoated MWCNTs surface after the Pt ALD process, which includes an ozone exposure step, the effect of the exposure of MWCNTs to ozone pulses used during the ALD deposition process on the oxidation temperature has to be taken into consideration. Therefore, an ozone treated reference sample of MWCNTs was achieved by applying 50 half-cycles of ozone pulse on MWCNTs with the same pulse time and same temperature as during the Pt ALD process. As can be seen in Fig. 6 (a), the variation of the carbon signal of the ozone treated reference sample overlaps with the uncoated MWCNTs, which excludes the possible effect of ozone treatment during the ALD process on the carbon removal. In Fig. 6 (b), the evolution of the grain size of Pt nanoparticles was calculated using the Sherrer equation and the FWHM of the Pt(111) diffraction peak during the calcination. The red points represent the evolution of the grain size during calcination in air at 450°C for 3 hours. The grain size starts to grow from 5nm, which is in a good agreement with the TEM characterization in Fig. 2, and then gradually increases as the annealing temperature is raised at a fixed rate of 5°C/min. When the temperature is above ~435°C, the grain size takes off and gradually increases during the isothermal plateau at 450°C and reaches a range of 24-27 nm at the end of the plateau. According to the EDX results in Fig. 6 (a), the carbon intensity drops steeply between 400 and 450°C. Since this coincides with the dramatic increase of the Pt grain size, the removal of the carbon appears to assist the merging of the Pt nanoparticles into larger grains. The limited increase of Pt grain size (blue points) of a control sample calcined at 450°C in He for 3 hours is in agreement with this interpretation. The grain size does not increase spectacularly when then sample was annealed at 400 °C in air for 3 hours (green points), although this long anneal also resulted in complete removal of the carbon according to EDX. This indicates that the calcination temperature is an important factor in the growth of the Pt grains.

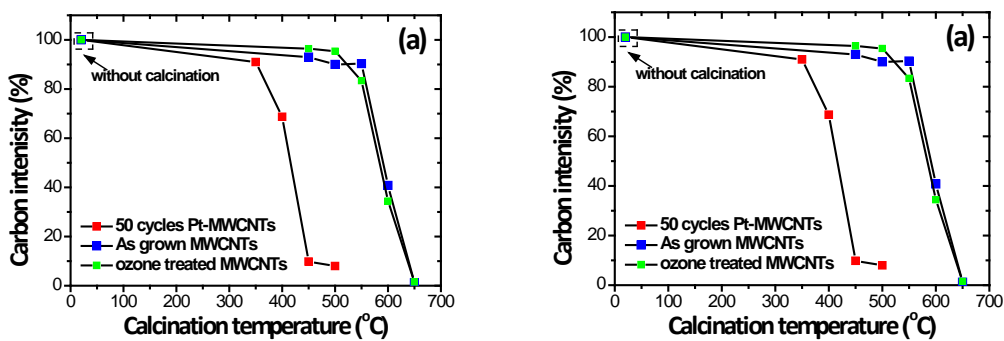

**Figure 6** (a) Carbon content as determined by EDS on as grown MWCNTs, ozone treated MWCNTs and 50 cycles Pt-MWCNTs calcined at different temperatures. (b) is the grain size versus calcination time, calculated from in situ XRD measurement by using the Sherrer equation. The sample was ramped up from room temperature to 400 or 450°C at a constant rate of 5°C/min, and then kept at this temperature for a period of up to 3 hours.

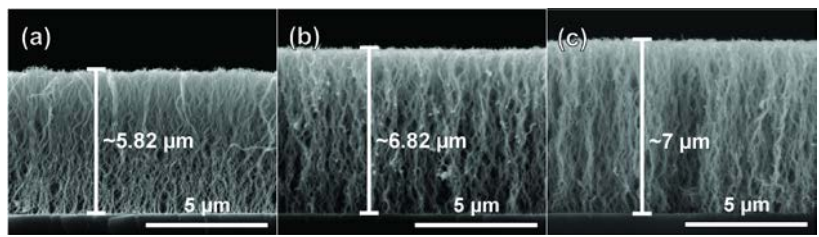

**Figure 7** SEM pictures of 50 cycles (a), 100 cycles (b,) and 200 cycles (c) Pt-WCNTs samples after calcination at 450°C in air for 3 hours.

To investigate the effect of the amount of Pt on the morphology after carbon removal, 100 and 200 cycles Pt-MWCNTs samples were also annealed using the same process. SEM images of the annealed samples are shown in Fig. 7(a) – (c). For all the samples, a similar morphology was observed for the 3D-Pt structure. However, a considerable decrease of the overall layer thickness was observed. The thicknesses of 50, 100 and 200 cycles 3D-Pt were measured as ~5.82, ~6.82 and ~7.0μm, respectively. The reduced shrinkage of the height for the sample with more ALD Pt cycles indicates that the presence of a larger amount of as deposited Pt coating reduces the shrinkage during calcination.

For the potential applications like water splitting and fuel cells, a high surface area enhancement of the 3D-Pt is a major advantage. To extract electrochemically active surface areas of both Pt-MWCNTs and 3D-Pt samples, electrochemical impedance spectroscopy (EIS) characterization was carried out. The surface area enhancement is determined by dividing electrochemical capacitance of Pt-MWCNTs and 3D-Pt samples with the capacitance of planar Pt sample. In Fig.8 (a) and (b), the CV measurements show that both Pt-MWCNTs and 3D-Pt samples exhibit high capacitance compared to the planar Pt reference sample, which is deposited by physically sputtering 100nm Pt and 70nm TiN on Si substrate. The surface area enhancement extracted by the EIS measurements on the 50, 100 and 200 cycles Pt-MWCNTs samples were 224, 145 and 106 $cm^2/cm^2$, respectively. For the 3D- Pt samples, surface area increase of 48, 47 and 54 $cm^2/cm^2$ for 50, 100 and 200 cycles samples were found, as shown by the solid symbols in Fig. 9. In order to extract the physically accessible surface area, we conformally deposited 50 ALD cycles ZnO into the synthesized sample and planar Pt sample using Diethylzinc (Sigmal Aldrich) and ozone. The ratio between Zn signal intensities of these two samples by XRF characterization allows obtaining the physically accessible surface area of the synthesized sample. Because of the non-uniform Pt coating on the MWCNTs, this method could only be applied to the 3D-Pt samples. As shown by the hollow triangle symbols in Fig. 9, the physically accessible surface areas of 50, 100, 200 3D-Pt samples are 45, 69 and 63 $cm^2/cm^2$, respectively.

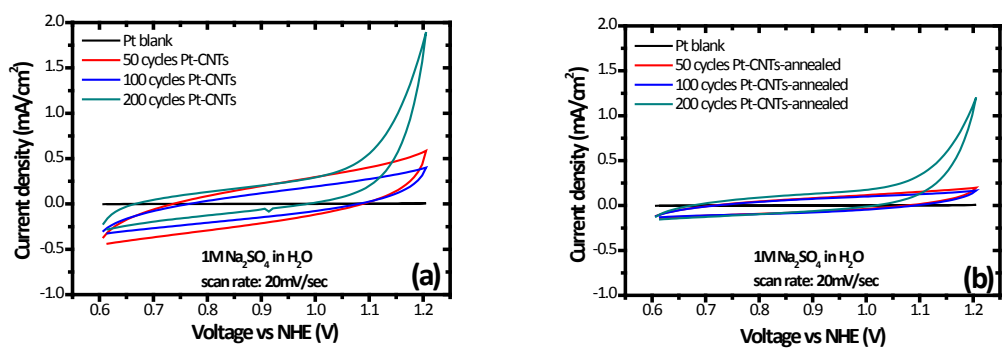

**Figure 8** Cyclic voltammograms of blank Pt substrate and Pt-MWCNTs (a) and 3D-Pt (b).

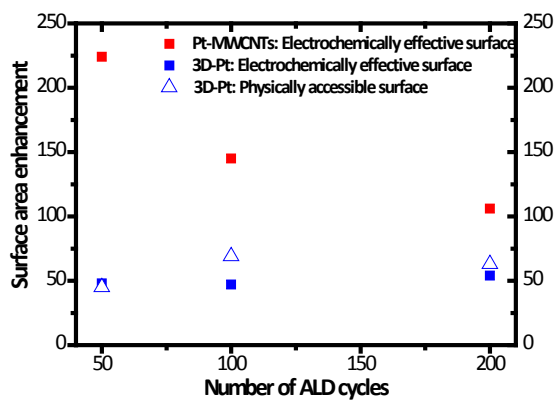

**Figure 9** Surface area enhancement of Pt-MWCNTs (red squares) and 3D-Pt porous structure extracted (blue squares) from EIS with potential 0.6V vs NHE and XRF (blue hollow triangles) measurements.

As shown in Fig. 9, the electrochemically effective surface area of the Pt-MWCNTs samples gradually decrease for Pt-MWCNTs as the number of ALD cycles changes from 50 to 200 cycles. This is attributed to the loss of porosity after mass loading of Pt into the sample. For the 3D Pt samples, both EIS measurement and XRF indicate that the electrochemical effective and physically accessible surface areas are independent of the number of ALD cycles, which might be due to the increase of the height as shown in Fig. 7. The good agreement of the physical and electrochemically effective surface area increases of the 3D-Pt samples indicates that the nearly entire physical area is electrochemically accessible. Compared to other methods of synthesizing Pt nanowires such as electrospinning, the method proposed in this work offers better potential for adhesion to the substrate and may offer a higher degree of controllability, e.g. by combining it with lithography to selectively form 3D-Pt structures onto specific regions of a sensor device. Even though the 3D-Pt only inherited a part of the surface area of the Pt-MWCNTs samples, the annealing process which induced nanowires instead of Pt nanoparticles coated on MWCNTs might show some benefits for applications which require the features of support-free, high temperature stability and resistance to corrosive environments.

## Conclusions

In summary, a novel porous 3D network comprised of free standing Pt nanowires was achieved by conformal ALD Pt on MWCNTs followed by annealing to remove the carbonaceous template. Quantitatively, an electrochemically active surface area enhancement of ~50 times was observed for the 3D-Pt sample. 2-D and 3-D TEM techniques applied to this network clearly show that Pt nanowires formed an interconnected architecture which consists of pure Pt without any carbon support. The porous, conductive and interconnected features of this 3D network of Pt nanowires could be ideal for multiple applications requiring support-free Pt catalysts, e.g. for use in corrosive environments.

## Acknowledgements

The authors wish to thank the Research Foundation - Flanders (FWO) for financial support. The authors acknowledge the European Research Council for funding under the European Union's Seventh Framework Programme (FP7/2007-2013)/ERCgrant agreement N°239865-COCOON, N°246791-COUNTATOMS and N°335078–COLOURATOM). The authors would also want to thank the support from

## References

1 F. E. Osterloh, *Chem. Mater.*, 2008, **20**, 35-54
2 S. Zhang, X. Z. Yuan, J. N.C. Hin, H. Wang, K. A. Friedrich and M .Schultz, *J. Power Sources*, 2009, **194**, 588
3 H. Tanaka, M. Taniguchi, M. Uenishi, N. Kajita, I. Tan, Y. Nishihata, J. Mizuki, K. Narita, M. Kimura and K. Kaneko, *Angew. Chem. Int. Ed.*, 2006, **45**, 5998-6002
4 J. Shui and J. C. M. Li, *Nano Lett.*, 2009, **9**, 1307-1314
5 D. Yang, S. Sun, H. Meng, J-P Dodelet and E. Sacher, *Chem. Mater.*, 2008, **20**, 4677-4681
6 S. Sun, G. Zhang, Y. Zhong, H. Liu, R. Li, X. Zhou and X. Sun, *Chem. Comm.*, 2009, 7048-7050
7 S. Sattayasamitsathit, A.M. O'Mahony, X. Xiao, S. M. Brozik, C. M. Washburn, D. R. Wheeler, J. Cha, D. B. Burckel, R. Polsky and J. Wang, *Electrochem. Comm.*, 2011, **13**, 856-860
8 Z. Shen, Y. Matsuki, K. Higashimine, M. Miyake and T. Shimoda, *Chem. Lett.*, 2012, **41**, 644-646
9 J. I. Shui, C. Chen and J. C. M. Li, *Adv. Fun. Mater.*, 2011, **21**, 3357-3362
10 X. Zhang, W. Lu, J. Da, H. Wang, D. Zhao and P. A. Webley, *Chem. Comm.*, 2009, 195-197
11 C. Detavernier, J. Dendooven, S. P. Sree, K. F. Ludwig, and J. A. Martens, *Chem. Soc. Rev.*, 2011, **40**, 5242.
12 J. Dendooven, B. Goris, K. Devloo-Casier, E. Levrau, E. Bier-mans, M. R. Baklanov, K. F. Ludwig, P. Van Der Voort, S. Bals, and C. Detavernier, *Chem. Mater.*, 2012, **24**, 1992-1994
13 D. J. Comstock, S. T. Christensen, J. W. Elam, M. J. Pellin and M. C. Hersam, *Adv. Funct. Mater.*, 2010, **20**, 3099-3105
14 Y. Zhou, D. M. King, X. Liang, J. Li and A. W. Weimer, *Appl. Catal. B: Environ.*, 2010, **101**, 54-60
15 J. A. Enterkin, K. R. Poeppelmeier and L. D. Marks, *Nano Lett.*, 2011, **11**, 993-997
16 C. Liu, C. C. Wang, C. C. Kei, Y. C. Hsueh and T. P. Perng, *Small*, 2009, **5**, 1535-1538
17 Z. Chen, M. Waje, W. Li and Y. Yan, *Angew. Chem. Int. Ed.*, 2007, **46**, 4060-4063
18 L. Liu and E. Pippel, *Angew. Chem. Int. Ed.*, 2011, **50**, 2729-2733
19 Q. Xie, Y.-L., Jiang, C. Detavernier, D. Deduytsche, R. L. Van Meirhaeghe, G.-P.Ru, B.-Z.Li, X.-P.Qu, *J. Appl. Phy.*, 2007, **102**, 083521
20 J. Dendooven, R. K. Ramachandran, K. Devloo-Casier, G. Ramelberg, M. Filez, H. Poelman, G. B. Marin, E. Fonda, and C. Detavernier, *J. Phys. Chem. C.*, 2013, **117**, 20557
21 J. Dendooven, S. PulinthanathuSree, K. De Keyser, D. Deduytsche, J. A. Martens, K. F. Ludwig, and C. Detavernier, *J. Phys. Chem. C*, 2011, **115**, 6605
22 X. Wang, S. M. Tabakman and H. Dai, *J. Am. Chem. Soc.*, 2008, **130**, 8152-8153
23 J. H. Lehman, M. Terrones, E. Mansfield, K. E. Hurst and V. Meunier, *Carbon*, 2011, **49**, 2581-2602